\documentclass[a4paper,10pt]{article}
\usepackage{ifthen}
\usepackage{axodraw}
\usepackage{epsfig}
\usepackage{graphicx}
\usepackage{array}
\usepackage{braket}

\usepackage{amsmath,amsthm,amssymb,amsfonts}
\usepackage{array}
\usepackage[center]{subfigure}
\usepackage{dsfont}
\usepackage{mathrsfs}
\usepackage{slashed}
\usepackage{epstopdf}
\newcommand{\ba}{\begin{eqnarray}}
\newcommand{\ea}{\end{eqnarray}}
\newcommand{\be}{\begin{equation}}
\newcommand{\ee}{\end{equation}}

\renewcommand{\arraystretch}{1.2}

\makeatletter
\def\fmslash{\@ifnextchar[{\fmsl@sh}{\fmsl@sh[0mu]}}
\def\fmsl@sh[#1]#2{%
  \mathchoice
    {\@fmsl@sh\displaystyle{#1}{#2}}%
    {\@fmsl@sh\textstyle{#1}{#2}}%
    {\@fmsl@sh\scriptstyle{#1}{#2}}%
    {\@fmsl@sh\scriptscriptstyle{#1}{#2}}}
\def\@fmsl@sh#1#2#3{\m@th\ooalign{$\hfil#1\mkern#2/\hfil$\crcr$#1#3$}}
\makeatother
\title{The decays $\Lambda_{b,c}\to N^*\, l\,\nu$ in QCD}
\author{\textbf{M. Emmerich, N. Offen, A. Sch\"afer}\\
Institut f\"ur theoretische Physik,\\ Universit\"at Regensburg, 93040 Regensburg, Germany}
\date{}

\begin{document}

\maketitle
\begin{abstract}
We present an exploratory study of the $\Lambda_{c,b}\to N^*$-form factors and the semileptonic decay width within the framework 
of light-cone sum rules. We use two different methods and two different interpolating currents for the $\Lambda_{c,b}$. 
\begin{enumerate}
 \item We follow \cite{Khodjamirian:2011jp} and eliminate negative parity partners of the $\Lambda_{c,b}$ by taking linear combinations of different 
 Lorentz-structures.
 \item We extract the form factors by choosing the Lorentz-structures with the highest possible powers of $p_+$.
\end{enumerate}
As interpolating currents we choose an axial-vector like and a pseudoscalar like current. Our results show that the procedure of eliminating negative parity partners 
is not well suited for the case at hand and that the second approach with an axial-vector like interpolating current gives the most reliable results.
Our predictions are based on the models obtained in \cite{Anikin:2015ita,Braun:2014wpa}. The largest uncertainty comes from 
the uncertainty of the twist 4 parameters $\eta_{10},\,\eta_{11}$ and we take the spread between the two models in \cite{Anikin:2015ita} 
as a measure for this. 
We get
\begin{eqnarray}\Gamma(\Lambda_b\to N^*(1535) l \nu)&=&\left(0.0058^{+0.0010}_{-0.0009}\right)\cdot\left(\dfrac{V_{ub}}{3.5\cdot10^{-3}}\right)^2,\quad\mbox{LCSR(1)} \nonumber\\
 \Gamma(\Lambda_b\to N^*(1535) l \nu)&=&\left(0.00070^{+0.00012}_{-0.00011}\right)\cdot\left(\dfrac{V_{ub}}{3.5\cdot10^{-3}}\right)^2,\quad\mbox{LCSR(2)} \nonumber\\
 \Gamma(\Lambda_c\to N^*(1535) l \nu)&=&\left(0.0064^{+0.0012}_{-0.0011}\right)\cdot \left(\dfrac{V_{cd}}{0.225}\right)^2,\quad\mbox{LCSR(1)}\nonumber\\
 \Gamma(\Lambda_c\to N^*(1535) l \nu)&=&\left(0.00077^{+0.00016}_{-0.00014}\right)\cdot \left(\dfrac{V_{cd}}{0.225}\right)^2,\quad\mbox{LCSR(2)}\nonumber
\end{eqnarray}
as predictions for the respective decay widths, where LCSR(1) and LCSR(2) refer to the two different models of the distribution amplitudes of the $N^*$. 
It is seen that even a rough measurement of these decays will greatly help to discriminate different models.
\end{abstract}

%%%%%%%%%%%%%%%%%%%%%%%%%%%%%%%%%%%%%%%%%%%%%%
\section{Introduction}
Measuring the properties of the nucleon resonances and interpreting them in terms of the fundamental degrees of freedom 
of QCD is one of the goals of the Hall B CLAS 12 detector at Jefferson Lab. \cite{Aznauryan:2012ba}
We propose here a complementary approach to nucleon resonances using the decays of $\Lambda_b$ and $\Lambda_c$ baryons 
produced abundantly at LHCb or at the planned PANDA experiment.\\
Our approach is based on light cone sum rules \cite{LCSR} a hybrid of classical 
SVZ-sum rules \cite{Shifman:1978bx} and methods from hard exclusive decays. It allows to relate the 
$\Lambda_{b,c}\to N^*$ decay form factors to the distribution amplitudes of the $N^*$ that is roughly speaking to the 
momentum distribution of the quarks in different Fock-states inside the $N^*$.\\
Sufficient experimental data will allow to compare the extracted distribution amplitudes to constraints 
coming from the lattice \cite{Braun:2009jy, Braun:2014wpa} or from electromagnetic $N^*$ form factors \cite{Anikin:2015ita}, see also 
\cite{Anikin:2013aka} for the general framework. This will provide an alternative approach to obtain information on the structure of 
low lying nucleon resonances, see also \cite{Oset:2016lyh}
for a review of hadronic decays of heavy mesons and baryons to investigate hadronic resonances.
Similar calculations for the $\Lambda_b \to N$ form factors have been done in \cite{Huang:2004vf, Wang:2009hra, Khodjamirian:2011jp, Azizi:2009wn}
and we will refer to \cite{Khodjamirian:2011jp} for more details on the calculation.\\
The paper is organized as follows: In section 2 we will give the definitions of the relevant form factors and a short introduction to the method 
of light cone sum rules. Section 3 is devoted to the numerical analysis of the sum rules and section 4 will give some concluding remarks.
The relevant formulas are given in the appendix. In addition we refer to \cite{Anikin:2015ita,Anikin:2013aka,Khodjamirian:2011jp}, 
see also \cite{Anikin:2013yoa, Anikin:2015qos}, for the basic definitions of the distribution amplitudes in order not to overload this publication.

%%%%%%%%%%%%%%%%%%%%%%%%%%%%%%%%%%%%%%%%%%%%%%%%%
\section{Form Factors and light cone sum rules}
We give the definitions and derivations for the case of the $\Lambda_c\to N^*$ decay. The transition to $\Lambda_b\to N^*$ is simply done by replacing $c\to b$ everywhere 
and $d\to u$ in the transition current.
The relevant form factors of the vector and axial-vector current are defined as
\begin{align}
&\bra{\Lambda_c(P')}j_\nu \ket{N^*(P)}&=\bar{u}_{\Lambda_c}(P') \left( f_1(q^2) \gamma_\nu + i\frac{ f_2(q^2)}{m_{\Lambda_c}}\sigma_{\nu \mu} q^\mu +\frac{f_3(q^2)}{m_{\Lambda_c}}q_\nu \right) \gamma_5 u_{N^*}(P),\nonumber\\
&\bra{\Lambda_c(P')}j_{\nu5} \ket{N^*(P)}&=\bar{u}_{\Lambda_c}(P') \left( g_1(q^2) \gamma_\nu + i\frac{ g_2(q^2)}{m_{\Lambda_c}}\sigma_{\nu \mu} q^\mu +\frac{g_3(q^2)}{m_{\Lambda_c}}q_\nu \right) u_{N^*}(P),
\label{eq:formfactor}
\end{align}
where $P'=P-q$. We don't consider $f_3$ and $g_3$ in the following since in semileptonic decays they will contribute with coefficients proportional to the lepton mass.\\
For the $\Lambda_c^*$-form factors one defines $\tilde{f}_i$ and $\tilde{g}_i$ analogously to (\ref{eq:formfactor}) by replacing $\Lambda_c$ with $\Lambda_c^*$ and adding a $\gamma_5$ 
 after the $\Lambda_c^*$-spinor. To get access to these form factors we use the correlation function
\begin{equation}
 \Pi_a(P,q)=i \int d^4x \hspace{2mm} e^{i q \cdot x} \bra{0}T \{\eta_{\Lambda_c}(0),j_a(x) \} \ket{N^*(P)},
\label{eq:corr}
 \end{equation}
where
 \[j_a(x)=\bar{c}\,\Gamma_a d(x),\quad \Gamma_a=\gamma_\nu,\,\gamma_\nu\gamma_5,\]
are the weak transition currents and
 \[\eta_{\Lambda_c}=\epsilon^{ijk}(u_i C\Gamma_1 d_j)\Gamma_2 c_k\]
is the interpolating current with the correct quantum numbers for the $\Lambda_c$. We will use two different choices for $\eta_{\Lambda_c}$ namely
\[\eta^{(P)}_{\Lambda_c}=(u\,C\gamma_5 d)c\quad\mbox{and}\quad \eta^{(A)}_{\Lambda_c}=(u\,C\gamma_5\gamma_\lambda d)\gamma^\lambda c.\]
The standard procedure of light cone sum rules is to calculate the correlation function (\ref{eq:corr}) in two different ways. On the one hand, one inserts a complete set 
of states with $\Lambda_c$ quantum numbers between the two currents and extracts the lowest lying state. On the other hand, one uses the operator product expansion (OPE) 
around the light cone for space like momenta $(P-q)^2,\,q^2\ll 0$. This results in two different representations of the correlation function which can be 
equated using dispersion relations and quark-hadron duality. Taking the Borel-transform to eliminate possible subtraction terms from the dispersion relations and 
to suppress higher states in the hadronic sum gives the final sum rule.\\

%%%%%%%%%%%%%%%%%%%%%%%%%%%%%%%%%%%%%%%%%%%%%%%%%%%
\subsection{Eliminating the $\Lambda_c^*$-pole}
We will give a few details on the procedure but refer the reader to \cite{Khodjamirian:2011jp} for all details of the calculations. 
To eliminate the $\Lambda_c^*$ pole from the sum rules, we explicitly keep both the $\Lambda_c$ and $\Lambda_c^*$ in the hadronic sum and represent higher states by a dispersion integral. 
The residue of the poles of the two states is given by a product of the form factors (\ref{eq:formfactor}) %changing $\Lambda_c$ to $\Lambda_c^*$ where necessary and adding $\gamma_5$ after the relevant spinor 
and their decay constants:
\begin{align}
& \bra{0}\eta_{\Lambda_c}^{(i)}\ket{\Lambda_c(P')}\,=\,\lambda_{\Lambda_c}^{(i)} m_{\Lambda_c} u_{\Lambda_c}(P'),\nonumber\\
&\bra{0}\eta_{\Lambda_c}^{(i)}\ket{\Lambda_c^*(P')}\,=\,\lambda_{\Lambda_c^*}^{(i)} m_{\Lambda_c^*} \gamma_5 u_{\Lambda_c^*}(P').
 \end{align}
Two main observations: First, using the equation of motion $(\fmslash{P}-m_{N^*})u_{N^*}(P)$, one can decompose the correlation function into six independent, invariant functions:
\begin{equation}
 \Pi_\mu^{(i)}(P,q)=\left(\widetilde{\Pi}_1^{(i)} P_\mu+\widetilde{\Pi}_2^{(i)} P_\mu\fmslash{q}+\widetilde{\Pi}_3^{(i)}\gamma_\mu+\widetilde{\Pi}_4^{(i)}\gamma_\mu\fmslash{q}+\widetilde{\Pi}_5^{(i)}q_\mu+
 \widetilde{\Pi}_6^{(i)}q_\mu\fmslash{q}\right)\gamma_5 u_{N^*}(P),
\label{eq:decompose}
 \end{equation}
where $i$ labels the pseudoscalar $P$ and axial-vector $A$ interpolating current.\\
Second, the form factors enter in front of more than one of the Lorentz-structures in (\ref{eq:decompose}). This allows, after one equates the hadronic sum and the OPE result, 
to construct linear combinations  where the contribution of the $\Lambda_c^*$ is eliminated. Taking e.g. the hadronic sum for the vector transition
\begin{align}
 &\Pi_\nu^{(i)}(P',q)= \frac{\lambda_{\Lambda_c}^{(i)} m_{\Lambda_c}}{m_{\Lambda_c}^2-P'^2} \left[ 2 f_1(q^2) P_\nu -2 \frac{f_2(q^2)}{m_{\Lambda_c}} P_\nu \slashed{q} \right. \nonumber \\
 &\left.+ (m_{N^*}+m_{\Lambda_c}) \left( f_1(q^2)+ \frac{m_{N^*}-m_{\Lambda_c}}{m_{\Lambda_c}} f_2(q^2) \right)\gamma_\nu + \left( f_1(q^2) +\frac{ m_{N^*}-m_{\Lambda_c}}{m_{\Lambda_c}} f_2(q^2) \right)\gamma_\nu \slashed{q} \right. \nonumber \\
 &\left.+\left( -2 f_1(q^2) -\frac{m_{N^*}-m_{\Lambda_c}}{m_{\Lambda_c}} (f_2(q^2)+f_3(q^2)) \right)q_\nu + \frac{1}{m_{\Lambda_c}} \left( f_2(q^2)-f_3(q^2)\right) q_\nu \slashed{q}  \right] \gamma_5 u_{N^*}(P) \nonumber \\
 &+ \frac{\lambda_{\Lambda_c^*}^{(i)} m_{\Lambda_c^*}}{m_{\Lambda_c^*}^2-P'^2} \left[ -2 \tilde{f}_1(q^2) P_\nu + 2 \frac{\tilde{f}_2(q^2)}{m_{\Lambda_c^*}} P_\nu \slashed{q} \right. \nonumber \\
 &\left. -(m_{N^*}-m_{\Lambda_c^*}) \left( \tilde{f}_1(q^2) +\frac{m_{N^*}+m_{\Lambda_c^*}}{m_{\Lambda_c^*}}\tilde{f}_2(q^2) \right) \gamma_\nu - \left( \tilde{f}_1(q^2)+ \frac{m_{N^*}+m_{\Lambda_c^*}}{m_{\Lambda_c^*}} \tilde{f}_2(q^2)\right)\gamma_\nu \slashed{q}\right.\nonumber \\
  &\left. +\left( 2 \tilde{f}_1(q^2) +\frac{m_{N^*}+m_{\Lambda_c^*}}{m_{\Lambda_c^*}} (\tilde{f}_2(q^2)+\tilde{f}_3(q^2)) \right) q_\nu  -\frac{1}{m_{\Lambda_c^*}} (\tilde{f}_2(q^2)-\tilde{f}_3(q^2)) q_\nu \slashed{q} \right] \gamma_5 u_{N^*}(P) \nonumber \\
 & +\int \limits_{s_0^h}^{\infty} \frac{ds}{s-P'^2} \bigg( \rho_1^{(i)}(s,q^2) P_\nu +\rho_2^{(i)}(s,q^2) P_\nu \slashed{q} \nonumber \\
 &+\rho_3^{(i)}(s,q^2) \gamma_\nu +\rho_4^{(i)}(s,q^2) \gamma_\nu \slashed{q} +\rho_5^{(i)}(s,q^2) q_\nu +\rho_6^{(i)}(s,q^2) q_\nu \slashed{q} \bigg) \gamma_5 u_{N^*}(P),
\label{eq:hadronsum1}
 \end{align}
where higher states are described by the spectral densities $\rho^{(i)}_j$ with $j=1,\,\ldots,\,6$. It can be seen that taking a linear combination of the first four Lorentz-structures 
one can get rid of $\tilde{f}_1(q^2),\,\tilde{f}_2(q^2)$ and get expressions for $f_1(q^2),\,f_2(q^2)$ in terms of $\Pi_1$ to $\Pi_4$.

%%%%%%%%%%%%%%%%%%%%%%%%%%%%%%%%%%%%%%%%%%%%%%%%%%%%
\subsection{Extracting highest powers of $p_+$}
We define a light-like vector $n_\mu$ by the condition
\begin{equation}
\label{n}
       q\cdot n =0\,,\qquad n^2 =0
\end{equation}
and introduce the second light-like vector as
\begin{equation}
\label{smallp}
p_\mu  = P_\mu  - \frac{1}{2} \, n_\mu \frac{m_N^2}{P\cdot n}\,,~~~~~ p^2=0\,,
\end{equation}
so that $P \to p$ in the infinite momentum frame, $P\cdot n\to \infty$, or
if the nucleon mass can be neglected, $m_N \to 0$.
The projector onto the directions orthogonal to $p$ and $n$ is then defined as
\begin{equation}
       g^\perp_{\mu\nu} = g_{\mu\nu} -\frac{1}{pn}(p_\mu n_\nu+ p_\nu n_\mu)\,.
\end{equation}
Taking a look at equations (\ref{eq:decompose}) and (\ref{eq:hadronsum1}) it is seen that contracting the correlation 
function $\Pi_\nu(P',q)$ with $n^\nu$ and multiplying by the projector $\frac{\fmslash{p}\fmslash{n}}{2p\cdot n}$ the result can be written as
\begin{equation}
 \dfrac{\fmslash{p}\fmslash{n}}{2p\cdot n} n^\nu\Pi_\nu(P',q)\,=\,p\cdot n\left(A(P',q)+\dfrac{\fmslash{q}_\perp}{m_{\Lambda_c}} B(P',q)\right).
\label{eq:funcAB}
 \end{equation}
The form factors $f_1(Q^2)$ and $f_2(Q^2)$ are then extracted from the sum rules for the functions $A(P',q)$ and $B(P',q)$ respectively.
$g_1(Q^2)$ and $g_2(Q^2)$ are extracted in the same way with the only difference being one additional $\gamma_5$.

%%%%%%%%%%%%%%%%%%%%%%%%%%%%%%%%%%%%%%%%%%%%%%%%%%%
\subsection{Deriving the light cone sum rules}%The operator product expansion}
For $P'^2,\,q^2\ll 0$ the product of two currents in (\ref{eq:corr}) can be expanded around the light-cone $x^2\sim 0$. 
At leading order the two $c$-quarks in (\ref{eq:corr}) are contracted giving the free propagator and the resulting matrix element is decomposed according to 
(A.21) in \cite{Anikin:2015ita}. This results in a sum over distribution amplitudes of different twists multiplied by their respective coefficient functions. Using equation of motions 
the contributions to the invariant functions $\widetilde{\Pi}_j$ in (\ref{eq:decompose}) can be identified. Neglecting terms which will vanish after Borel-transformation they can in general be written as 
\begin{equation}
 \widetilde{\Pi}_j^{(i)}(P'^2,q^2)=\frac{1}{4}\sum_{n=1,2,3} \int \limits_{0}^{1} dx \frac{w_{jn}^{(i)}(x,q^2)}{D^n},
 \label{eq:Pi-general}
\end{equation}
with the denominator
\begin{equation}
  D=m_c^2-(xP-q)^2=m_c^2-xP'^2-\bar{x}q^2+x \bar{x} m_{N^*}^2 ,
 \label{eq:denom}
 \end{equation}
where $\bar{x}=1-x$. The different functions $w_{jn}^{(i)}$ are distinguished by their indizies, where $i$ denotes either $A$ or $P$ for axial- or pseudoscalar interpolating current, 
$j=1,\,\ldots,6$ parametrizes the invariant amplitude it contributes to and $n=1,\,\ldots,\,3$ is the power of the denominator. The functions $w_{jn}^{(i)}$ are given in appendix \ref{functions}.\\
We have to write (\ref{eq:Pi-general}) as a dispersion integral in $P'^2$:
\begin{equation}
 \widetilde{\Pi}_j^{(i)}(P'^2,q^2)=\dfrac{1}{\pi}\int_{m_c^2}^\infty \dfrac{ds}{s-P'^2}\,\mbox{Im}_s\,\widetilde{\Pi}_j^{(i)}(s,q^2).
\end{equation}
Therefore we substitute
\begin{eqnarray}
 s(x)&=&\dfrac{1}{x}(m_c^2-\bar{x}q^2+x\bar{x}m_{N^*}^2),\nonumber\\
 x(s)&=&\dfrac{1}{2m_{N^*}^2}\left[m_{N^*}^2+q^2-s+\sqrt{(s-q^2-m_{N^*}^2)^2+4m_{N^*}^2(m_c^2-q^2)}\right]
\end{eqnarray}
in the denominator (\ref{eq:denom}) and do a partial integration if the power of the denominator is larger than one. Using quark-hadron duality to approximate 
the contributions of the hadronic states
\begin{equation}
 \int_{s_0^h}^\infty\dfrac{ds}{s-P'^2}\rho^{(i)}_j(q^2)\approx\dfrac{1}{\pi}\int_{s_0}^\infty\dfrac{ds}{s-P'^2}\,\mbox{Im}_s\,\widetilde{\Pi}^{(i)}_j(q^2),
\end{equation}
where $s_0$ is the duality threshold, and performing a Borel-transformation that results in
\[\dfrac{1}{s-P'^2}\to e^{-s/M^2},\quad (P'^2)^n\to0,\]
with $M^2$ being the Borel-parameter leads to the final sum rules
\begin{align}
 & f_1(q^2)= \frac{e^{m_{\Lambda_c}^2/M^2}}{2 m_{\Lambda_c}(m_{\Lambda_c}+m_{\Lambda_c^*}) \lambda_{\Lambda_c}^{(i)}} \frac{1}{\pi} \int \limits_{m_c^2}^{s_0} ds \hspace{1mm} e^{-s/M^2} \left[(m_{\Lambda_c}-m_{N^*}) \left( \text{Im}_s \tilde{\Pi}_1^{(i)} (s,q^2)  \right. \right. \nonumber \\
  &\left. \left.  -(m_{\Lambda_c^*}+m_{N^*}) \text{Im}_s\tilde{\Pi}^{(i)}_2(s,q^2)\right) +2 \text{Im}_s \tilde{\Pi}^{(i)}_3(s,q^2)+2 (m_{\Lambda_c^*}-m_{\Lambda_c}) \text{Im}_s \tilde{\Pi}^{(i)}_4(s,q^2) \right] ,\nonumber\\
  &f_2(q^2)= \frac{e^{m_{\Lambda_c}^2/M^2}}{2(m_{\Lambda_c}+m_{\Lambda_c^*})\lambda_{\Lambda_c}^{(i)} } \frac{1}{\pi} \int \limits_{m_c^2}^{s_0} ds \hspace{1mm} e^{-s/M^2} \left[ \text{Im}_s \tilde{\Pi}^{(i)}_1(s,q^2) \right. \nonumber \\
 &\left. -(m_{\Lambda_c^*}+m_{N^*}) \text{Im}_s \tilde{\Pi}^{(i)}_2(s,q^2)-2 \text{Im}_s \tilde{\Pi}^{(i)}_4(s,q^2)  \right],
\end{align}
with subtraction of the $\Lambda_c^*$-pole and 
\begin{align}
 &f_1(q^2)= \frac{e^{m_{\Lambda_c}^2/M^2}}{2 m_{\Lambda_c} \lambda_{\Lambda_c}^{(i)}} \frac{1}{\pi} \int \limits_{m_c^2}^{s_0} ds \hspace{1mm} e^{-s/M^2}  \text{Im}_s \tilde{\Pi}_1^{(i)} (s,q^2),\nonumber\\
 &f_2(q^2)= \frac{e^{m_{\Lambda_c}^2/M^2}}{2 \lambda_{\Lambda_c}^{(i)}} \frac{1}{\pi} \int \limits_{m_c^2}^{s_0} ds \hspace{1mm} e^{-s/M^2}  \text{Im}_s \tilde{\Pi}_2^{(i)} (s,q^2),
 \end{align}
for those without subtraction. Again the procedure for $g_1$ and $g_2$ is the same with some trivial sign differences due to an additional $\gamma_5$.
%Since we want to bring this in the form of a dispersion integral
%\[\widetilde{\Pi}_j^{(i)}(P'^2,q^2)=\int \dfrac{ds}{s-P'^2}\,\mbox{Im}_s\widetilde{\Pi}_j^{(i)}(s,q^2)\]
%we substitute 

%%%%%%%%%%%%%%%%%%%%%%%%%%%%%%%%%%%%%%%%%%%%%%%%%%%%%
\section{Numerical Analysis}
\subsection{Prerequisites}
We start this section by specifying the input parameters we use for our numerical analysis. The masses of the involved Baryons are taken from \cite{Agashe:2014kda} 
and for the $\Lambda_b^*$ from \cite{Wang:2010fq}
\begin{align}
&m_{\Lambda_c}=2.286\, \mbox{GeV}, & m_{\Lambda_c^*}=2.595\, \mbox{GeV},\nonumber\\
&m_{\Lambda_b}=5.620\,\mbox{GeV},& m_{\Lambda_b^*}=5.85\, \mbox{GeV}. \,\,\;\;
\end{align}
The values for the shape parameters of the $N^*$-distribution amplitude are given in table 1. For the twist 4 normalization factors
$\lambda_1^{(A,P)}$ we use the corresponding leading order two point sum rules instead of inserting fixed values. We take the same two approaches to the two point sum rules 
as for the light-cone sum rules. 
The correlation functions for $\lambda_1^{(A,P)}$ and $\lambda_1^{(A,P)}\cdot f_i^{(A,P)}$ have a very similar 
$\mu$-dependence which effectively cancels in the quotient. This has the advantage of considerably 
reducing the $\mu$-dependence of the result. We take the sum rules from \cite{Bagan:1993ii}, see also \cite{Khodjamirian:2011jp}\footnote{Note a typo in the dimension six part of Im $\tilde{F}_1(s)$, eq. (91), in \cite{Khodjamirian:2011jp}.
The correct expression is \cite{Yuming}
\[\mbox{Im}\tilde{F}_1^{dim\,6}(s)=\pi\dfrac{\langle\bar{q} q\rangle^2}{72}\delta(s-m_c^2)(11+2b-13b^2)\]} 
and have checked their results.

\begin{table*}[t]
\renewcommand{\arraystretch}{1.2}
\begin{center}
\scriptsize
\begin{tabular}{@{}l|l|l|l|l|l|l|l|l|l|l@{}} \hline
Method     & $\vert\lambda^{N^\ast}_1/\lambda^{N}_1\vert $ &$f_{N^\ast}/\lambda^{N^\ast}_1 $ & $\varphi_{10}$ & $\varphi_{11}$ & $\varphi_{20}$ & $\varphi_{21}$ & $\varphi_{22}$ & $\eta_{10}$   & $\eta_{11}$    & Ref. \\ \hline
LCSR (1) & 0.633     &0.027      & 0.36     & -0.95      & 0       & 0        & 0         & 0.00    & 0.94   & \cite{Anikin:2015ita} \\ \hline
LCSR (2) & 0.633     &0.027      & 0.37     & -0.96      & 0       & 0        & 0         & -0.29   & 0.23   & \cite{Anikin:2015ita} \\ \hline
LATTICE  & 0.633(43) &0.027(2)   & 0.28(12)  & -0.86(10)  & 1.7(14) & -2.0(18) & 1.7(26)   & -       & -      & \cite{Braun:2014wpa} \\ \hline
\end{tabular}
\end{center}

\caption[]{\small\sf Parameters of the $N^\ast(1535)$ distribution amplitudes at the scale $\mu^2=2$~GeV$^2$.
              For the lattice results \cite{Braun:2014wpa} only statistical errors are shown.
The set of parameters indicated as LCSR~(1) corresponds to the fit to the form factors $G_1(Q^2)$ and $G_2(Q^2)$
extracted from the measurements of helicity amplitudes in Ref.~\cite{Aznauryan:2009mx}
adding the errors in quadrature. The set of parameters indicated as LCSR~(2) is obtained from the fit to helicity amplitudes including
all available data at $Q^2 \ge 1.7~\text{GeV}^2$~\cite{Denizli:2007tq,Dalton:2008aa,Armstrong:1998wg,Aznauryan:2009mx}.
 $\lambda_1^N$ is given in \cite{Braun:2014wpa} as $10^2\,m_N\,\lambda_1^N=-3.88(2)(19)$ GeV$^3$.Note the typo in the first column in \cite{Anikin:2015ita}.\label{table1}}
 \end{table*}
 A thorough analysis of the sum rules shows that those derived by eliminating the $\Lambda_{c,b}^*$-pole are plagued by several problems:
 \begin{enumerate}
  \item Except for the form factors $f_1(Q^2)$ and $g_1(Q^2)$ using the pseudoscalar interpolating current there are large numerical cancellations 
  between different Lorentz-structures
  \item For the pseudoscalar current there is generally very little hierarchy between contributions of different twists, there are numerical cancellations 
  between different twists and they are very sensitive to the variation of the higher twist parameter $\xi_{10}$
  \item There is no set of parameters $M^2,\,s_0$ so that the sum rules for all four form factors fullfill the basic criteria used to check their viability  
 \end{enumerate}
Finally we have chosen the sum rules for the functions $A$ and $B$, see (\ref{eq:funcAB}), with the axial vector interpolating current and the respective two-point sum rule with 
the highest powers of $p_+$ as our default. The two models from table 1 give a measure for the uncertainty coming from the variation of twist 4 parameters $\eta_{10}$ and $\eta_{11}$. 
We quote the results for the two models separately to demonstrate the ability to discern different models by measuring theses decays.
The pseudoscalar interpolating current has similar problems as mentioned above. In most channels there is no clear hierarchy between different twists and there occur 
large numerical cancellations. In addition the sum rules for the pseudoscalar current are very sensitive to the basically unknown parameter $\xi_{10}$. Varying this 
parameter in the range of $-0.2\leq\xi_{10}\leq0.2$ changes the result by up to a factor of 6. In contrast the axial-vector sum rules depend only very mildly on this parameter.
 The Borel-parameter and duality threshold are chosen in a way that the usual sum rule criteria, that is the suppression of continuum states and of higher twist contributions, are fullfilled.
 \begin{figure}[h]
 \begin{center}
  \includegraphics[width=0.49\textwidth]{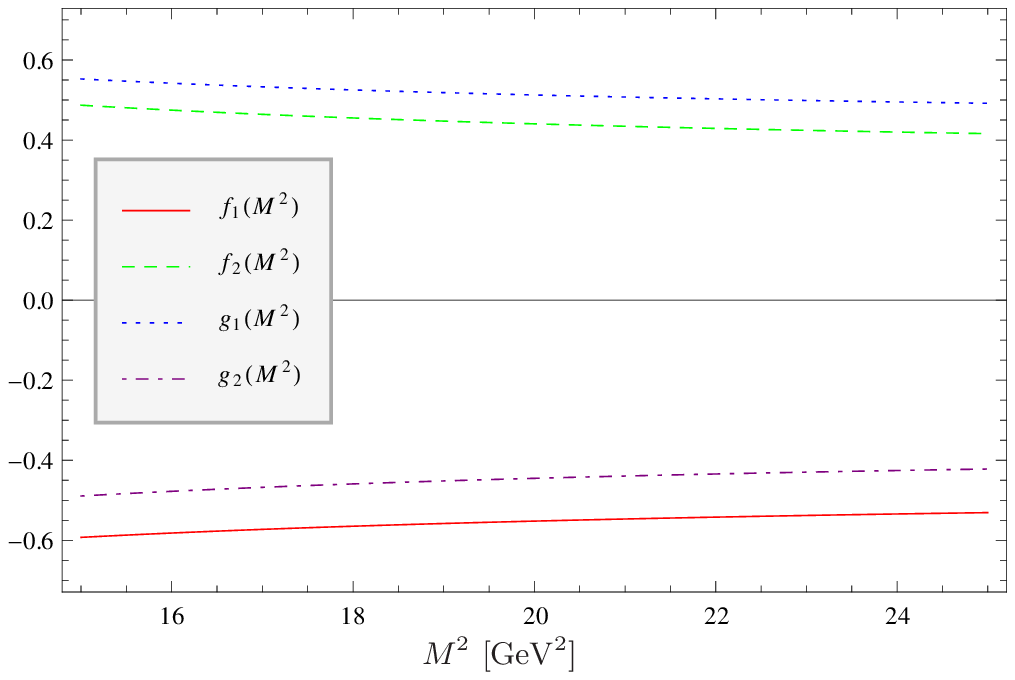}\hspace{0.2cm}\includegraphics[width=0.49\textwidth]{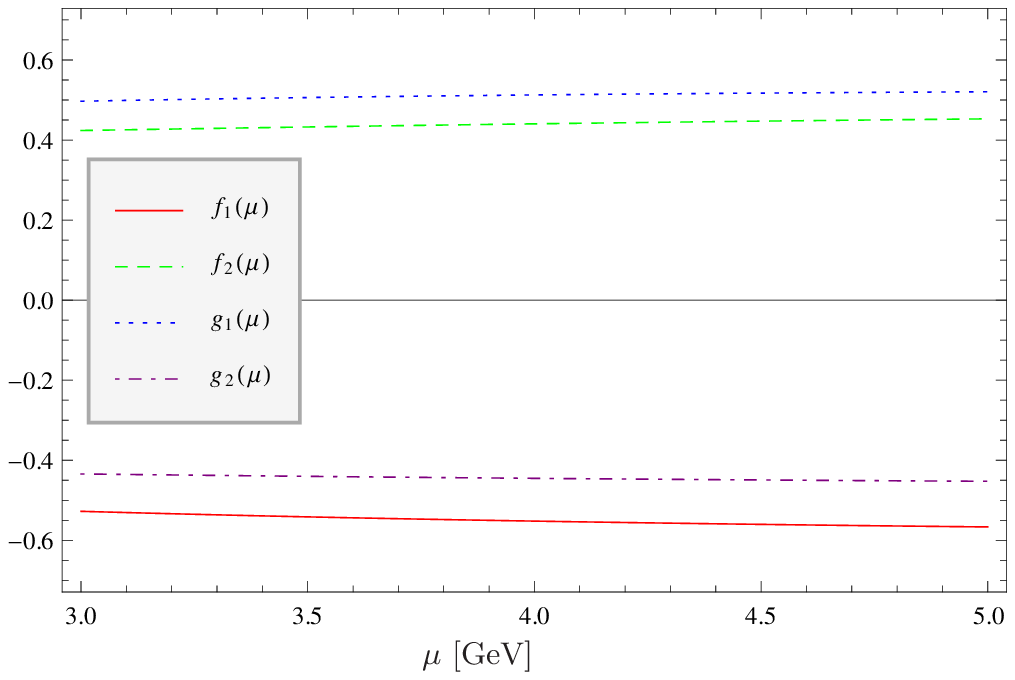}
 \end{center}
 
 \caption{The Borel- and $\mu$-dependence of the $\Lambda_b\to N^*$-form factors at $Q^2=0$\label{figure1}}
 \end{figure}
 We observe that for the range of 
 \begin{eqnarray}
 M_b^2=15-25\,\mbox{GeV}^2&\quad& s^b_0=36-40\,\mbox{GeV}^2\nonumber\\
 M_c^2=5-10\,\mbox{GeV}^2&\quad& s^c_0=6-10\,\mbox{GeV}^2\nonumber
 \end{eqnarray}
 these criteria are fullfilled and 
 that the sum rules are reasonably stable with respect to the variation of the Borel-parameter, see figure 1.\\ 
\subsection{Decay-widths}
Since the light cone sum rules are only valid up to $q^2\leq q_{max}^2=(m_{\Lambda_{c,b}}-m_{N^*}^2)^2$ we need to extrapolate the 
sum rule results to the whole physical region to calculate the decay-widths. We do this by making a two-parameter fit to our numerical values using the fit-function proposed 
in \cite{Bourrely:2008za}
\begin{eqnarray}
 f_i(q^2)&=&\dfrac{f_i(0)}{1-\frac{q^2}{m_{B^*(1^-)}^2}}\left\{1+b_i\left(z(q^2,t_0)-z(0,t_0)\right)\right\}\nonumber\\
 g_i(q^2)&=&\dfrac{g_i(0)}{1-\frac{q^2}{m_{B^*(1^+)}^2}}\left\{1+\tilde{b}_i\left(z(q^2,t_0)-z(0,t_0)\right)\right\}
 \label{eq:fitformula}
\end{eqnarray}
with
\begin{equation}
 z(q^2,t_0)=\dfrac{\sqrt{t_+-q^2}-\sqrt{t_+-t_0}}{\sqrt{t_+-q^2}+\sqrt{t_+-t_0}}
\end{equation}
and
\begin{eqnarray}
 t_\pm&=&(m_{\Lambda_b}\pm m_{N^*})^2,\nonumber\\
 t_0&=&t_+-\sqrt{t_+-t_-}\sqrt{t_+-t_{min}}.
\end{eqnarray}
$t_{min}$ is the lowest value of $q^2$ mapped to $z(q^2,t_0)$, so that $t_{min}=q^2_{min}\leq q^2\leq t_-$.  
We extend the fit region by calculating the form factor starting from $q^2_{min}=-6$ GeV$^2$.
%especially for the $\Lambda_c\to N^*$-decay we have calculated the form 
%factors starting from $q^2=-6$ GeV$^2$. 
We perform a weighed fit using as weights the uncertainties coming from the input parameters of the sum rules added in quadrature. For asymmetric errors we take the mean 
value and shift the central value by the difference of the mean value and the asymmetric error to get symmetric errors.
The fits compared to our sum rule values can be found in figure 2 and the results in table 2.\\
\begin{figure}[h]
 \begin{center}
  \epsfig{file=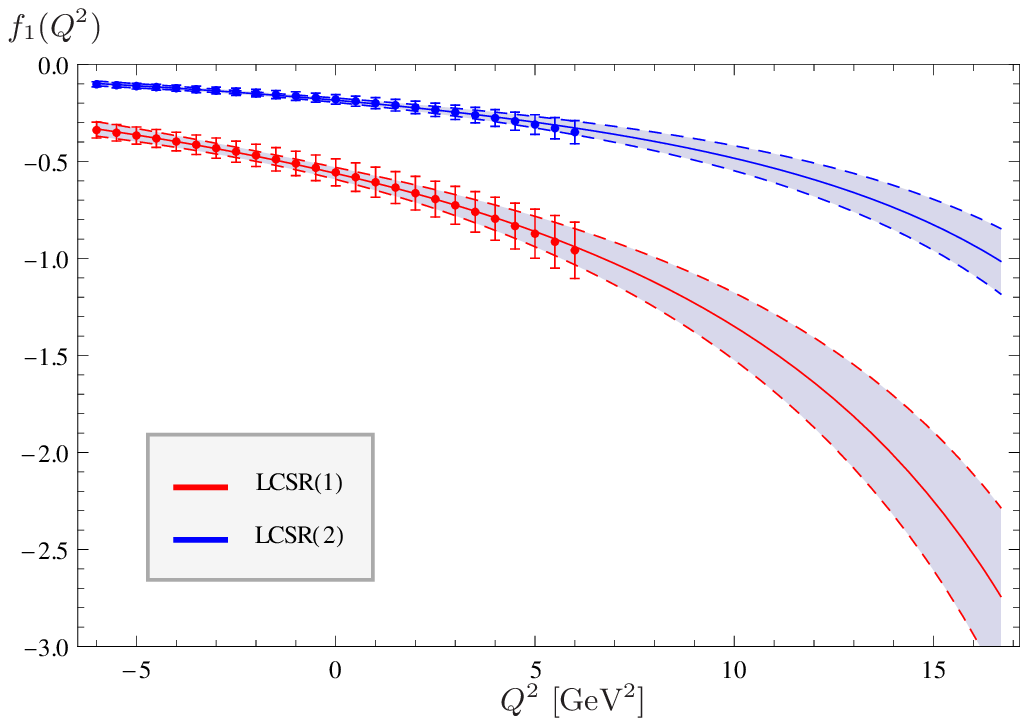, scale=0.65}\hspace{0.2cm}\epsfig{file=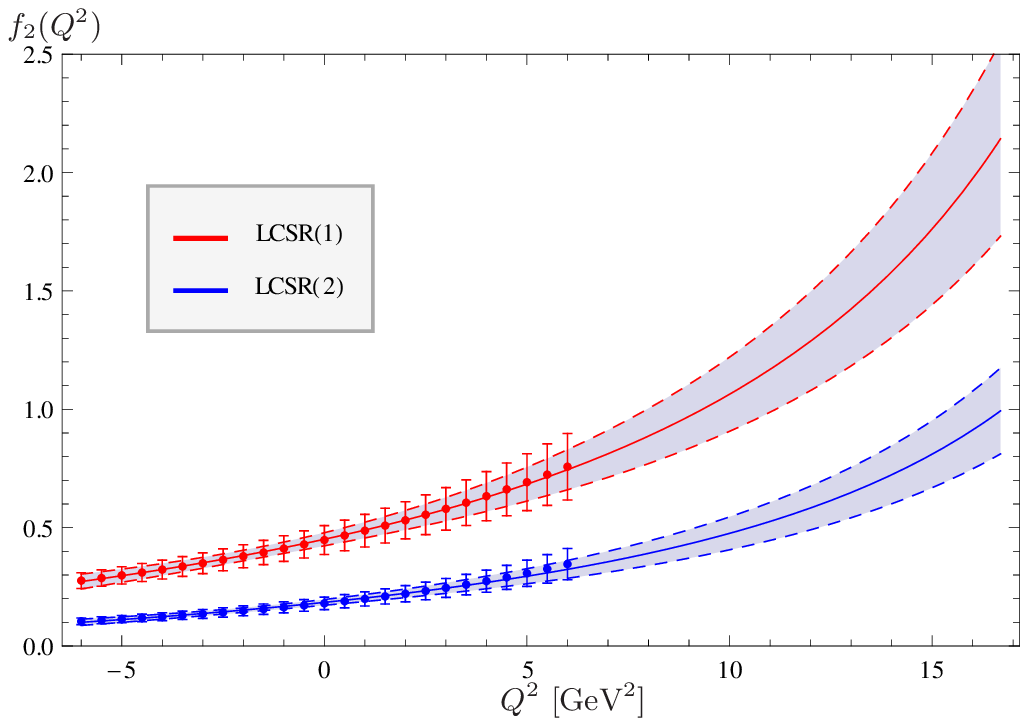, scale=0.65}
  \epsfig{file=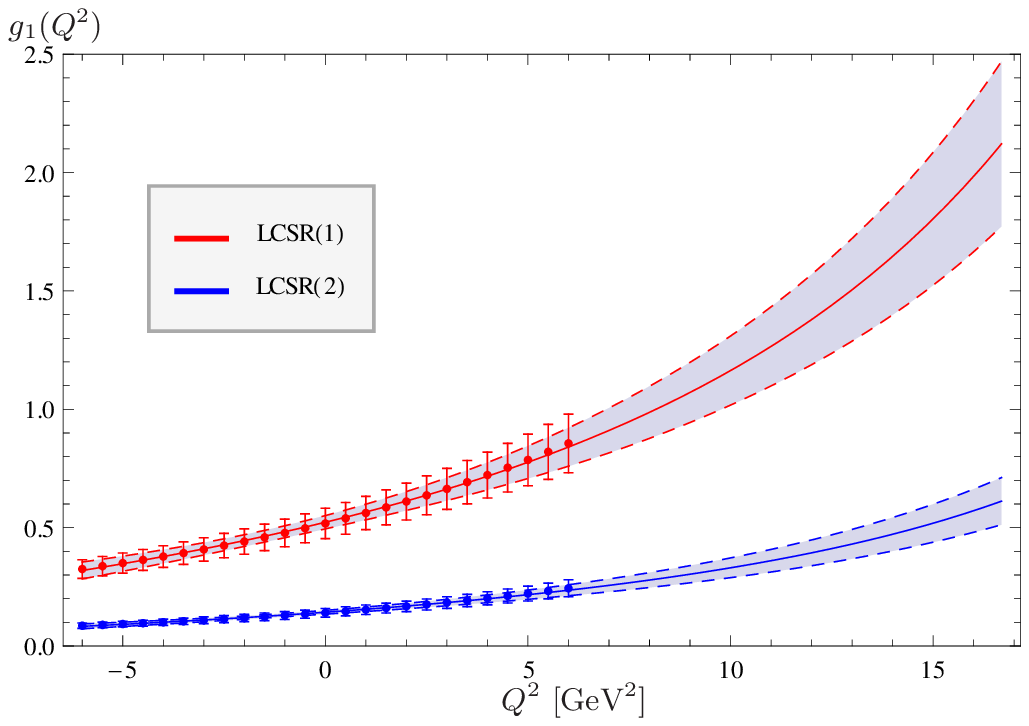, scale=0.65}\hspace{0.2cm}\epsfig{file=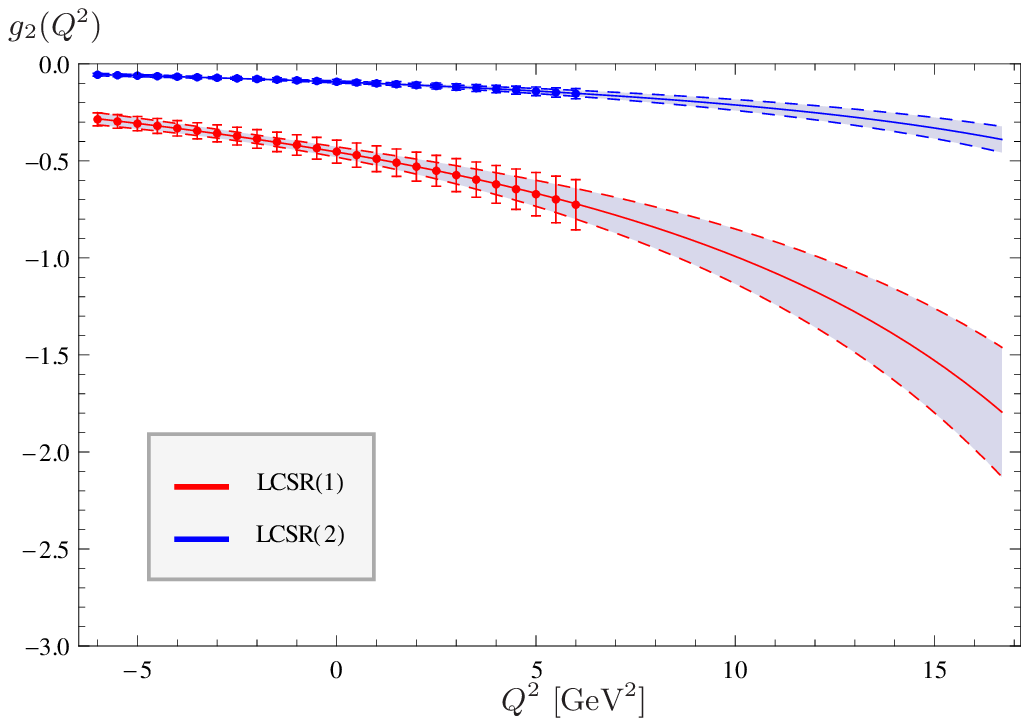, scale=0.65}

\caption{The four different form factors for the decay $\Lambda_b\to N^*$. LCSR(1) and LCSR(2) refer to the two models from table 1. The thick lines give the central value of the fits 
to the sum rule results (large dots), the dashed lines give the 2 $\sigma$ uncertainties coming from all parameters except twist 4.\label{Fig2} }
\end{center}
\end{figure}
\begin{table}[h]
 \begin{tabular}{|c|c|c|c|c|c|}
  \multicolumn{6}{c}{\rule[-3mm]{0mm}{8mm}\textbf{$\Lambda_b\to N^*$}}\\\hline\hline
  \multicolumn{3}{|c|}{LCSR(1)}&\multicolumn{3}{|c|}{LCSR(2)}\\\hline
  fit parameter& estimate & uncertainty&fit parameter& estimate & uncertainty\\\hline
  $f_1(0)$&-0.562&0.015&$f_1(0)$&-0.185&0.005\\\hline
  $b_1$&-10.236&1.420& $b_1$&-12.803&1.505\\\hline
 $f_2(0)$&0.451&0.0133& $f_2(0)$&0.184&0.006\\\hline
 $b_2(0)$&-9.695&1.549& $b_2(0)$ &-12.355&1.623\\\hline
 $g_1(0)$&0.523&0.014& $g_1(0)$ &0.143&0.004\\\hline
 $\tilde{b}_1$&-10.050&1.401&$\tilde{b}_1$&-11.205&1.460\\\hline
 $g_2(0)$&-0.454&0.013&$g_2(0)$&-0.093&0.003\\\hline
 $\tilde{b}_2$&-9.521&1.500&$\tilde{b}_2$&-10.718&1.483\\\hline
 \multicolumn{6}{c}{\rule[-3mm]{0mm}{8mm}\textbf{$\Lambda_c\to N^*$}}\\\hline\hline
  \multicolumn{3}{|c|}{LCSR(1)}&\multicolumn{3}{|c|}{LCSR(2)}\\\hline
  fit parameter& estimate & uncertainty&fit parameter& estimate & uncertainty\\\hline
  $f_1(0)$&-2.17&0.18&$f_1(0)$&-0.74&0.07\\\hline
  $b_1$&-6.26&1.06& $b_1$&-7.44&0.89\\\hline
 $f_2(0)$&1.09&0.12& $f_2(0)$&0.61&0.06\\\hline
 $b_2(0)$&-5.91&1.01& $b_2(0)$ &-7.14&0.76\\\hline
 $g_1(0)$&1.70&0.14& $g_1(0)$ &0.53&0.05\\\hline
 $\tilde{b}_1$&-6.65&1.01&$\tilde{b}_1$&-7.18&1.01\\\hline
 $g_2(0)$&-0.98&0.11&$g_2(0)$&-0.10&0.03\\\hline
 $\tilde{b}_2$&-6.44&0.96&$\tilde{b}_2$&-7.35&2.29\\\hline
 \end{tabular}
\caption{The fit parameters according to equation (\ref{eq:fitformula}) for the $\Lambda_{b,c}\to N^*$ form factors and the two models LCSR(1) and LCSR(2)}
\end{table}
As can be seen the largest uncertainty comes from the twist 4 parameters which are taken from table 1. This means these decays are very 
sensitive to the shape of the distribution amplitudes of the $N^*$ that parametrize relative orbital angular momentum of the quarks. 
We calculate the decay width by the following expression:
 \begin{eqnarray}
 \frac{d\Gamma}{dq^2}(\Lambda_b\to p l \nu_l) = {G_F^2
 m_{\Lambda_b}^3 \over 192 \pi^3 } |V_{ub}|^2 \lambda^{1/2}(1, r^2,
t)  \bigg \{ [(1-r)^2-t][(1+r)^2+2 t] |g_1(q^2)|^2 && \nonumber \\
  + [(1+r)^2-t][(1-r)^2+2 t] |f_1(q^2)|^2
-6 t [(1-r)^2 - t]  (1+r) g_1(q^2) g_2(q^2) && \nonumber \\
 -6 t [(1+r)^2 - t]  (1-r) f_1(q^2) f_2(q^2) + t [(1-r)^2-t] [2 (1+r)^2+t] |g_2(q^2)|^2  && \nonumber \\
  + t [(1+r)^2-t] [2 (1-r)^2+t] |f_2(q^2)|^2\bigg \}  \,. &&
\label{eq:decaywidth}
  \end{eqnarray}
Compared to \cite{Khodjamirian:2011jp, Azizi:2009wn} we have to exchange $f_1\leftrightarrow g_1$ and $f_2\leftrightarrow g_2$ and we neglect $f_3$ and $g_3$ 
since their coefficient is always proportional to the lepton mass. Our results are
\begin{eqnarray}
 \Gamma(\Lambda_b\to N^*(1535) l \nu)&=&\left(0.0058^{+0.0010}_{-0.0009}\right)\cdot\left(\dfrac{V_{ub}}{3.5\cdot10^{-3}}\right)^2,\quad\mbox{LCSR(1)} \nonumber\\
 \Gamma(\Lambda_b\to N^*(1535) l \nu)&=&\left(0.00070^{+0.00012}_{-0.00011}\right)\cdot\left(\dfrac{V_{ub}}{3.5\cdot10^{-3}}\right)^2,\quad\mbox{LCSR(2)} \nonumber\\
 \Gamma(\Lambda_c\to N^*(1535) l \nu)&=&\left(0.0064^{+0.0012}_{-0.0011}\right)\cdot \left(\dfrac{V_{cd}}{0.225}\right)^2,\quad\mbox{LCSR(1)}\nonumber\\
 \Gamma(\Lambda_c\to N^*(1535) l \nu)&=&\left(0.00077^{+0.00016}_{-0.00014}\right)\cdot \left(\dfrac{V_{cd}}{0.225}\right)^2,\quad\mbox{LCSR(2)}\nonumber
\end{eqnarray}
where we have given the results for the two different models separately to support our claim that the measurement of these decay widths will help to understand the structure of 
the twist 4 distribution amplitudes of the $N^*$.
%the largest part of the uncertainty comes due to the uncertainty in the twist 4 parameters. We have taken the two models 
%from table 1 as a measure for the range of values $\eta_{10}$ and $\eta_{11}$ can take. 
 %%%%%%%%%%%%%%%%%%%%%%%%%%%%%%%%%%%%%%%%%%%%%%%%%
\section{Conclusions}
We propose to measure the $\Lambda_{b,c}\to N^*$ form factors at PANDA and LHCb as an alternate way to extract information on the distribution amplitudes of the $N^*$. 
We did a leading order calculation in the framework of light cone sum rules taking into account three particle Fock-states 
up to twist 6.\\
We eliminated the $\Lambda_{b,c}^*$ contribution following \cite{Khodjamirian:2011jp} and investigated two different interpolating currents for 
the $\Lambda_{b,c}$. We found that all our sum rules using the elimination of the $\Lambda_{b,c}^*$-pole are plagued by large numerical cancellations between 
different Lorentz-structures except for the form factors $F_1(Q^2)$ and $G_1(Q^2)$ if one uses
the pseudoscalar interpolating current. While this might look like a good sign these sum rules have the problem that there 
are large cancellations between contributions of different twist. Our final verdict is, that the elimination of $\Lambda_{b,c}^*$ is not well suited for the case at hand: 
The mixing of different Lorentz-structures leads to a larger continuum contribution and to large numerical cancellations. 
The only two exceptions where the method could work are sum rules which have bad characteristics for each Lorentz-structure separately.\\
The comparison of the two interpolating currents leads to the conclusion that the axial-vector current is generally better suited 
for our calculation. It shows a clearer hirarchy of contributions of different twist, lesser cancellations and a lower sensitivity to the only badly known
$\lambda_2^{N^*}$ and $\xi_{10}$.\\ 
In general it is seen that in contrast to the $\Lambda_{b,c}\to N$ case the sum rules are dominated by twist 4 contribution, 
i.e. contributions with one unit of relative angular momentum, due to the higher mass of the $N^*$ and the smaller 
normalization factor of the leading order distribution amplitude $f_{N^*}$. A similar feature although not quite as pronounced was already observed in \cite{Anikin:2015ita}.\\
This circumstance makes this decay an ideal candidate to constrain the twist 4 distribution amplitudes.
The predicted decay widths for the two models LCSR(1) and LCSR(2) 
\begin{eqnarray}
 \Gamma(\Lambda_b\to N^*(1535) l \nu)&=&\left(0.0058^{+0.0010}_{-0.0009}\right)\cdot\left(\dfrac{V_{ub}}{3.5\cdot10^{-3}}\right)^2,\quad\mbox{LCSR(1)}, \nonumber\\
 \Gamma(\Lambda_b\to N^*(1535) l \nu)&=&\left(0.00070^{+0.00012}_{-0.00011}\right)\cdot\left(\dfrac{V_{ub}}{3.5\cdot10^{-3}}\right)^2,\quad\mbox{LCSR(2)}, \nonumber\\
 \Gamma(\Lambda_c\to N^*(1535) l \nu)&=&\left(0.0064^{+0.0012}_{-0.0011}\right)\cdot \left(\dfrac{V_{cd}}{0.225}\right)^2,\quad\mbox{LCSR(1)},\nonumber\\
 \Gamma(\Lambda_c\to N^*(1535) l \nu)&=&\left(0.00077^{+0.00016}_{-0.00014}\right)\cdot \left(\dfrac{V_{cd}}{0.225}\right)^2,\quad\mbox{LCSR(2)},\nonumber
\end{eqnarray}
support this claim by showing that even a rough measurement would already be able to discriminate between the two models.
With upcoming experimental data and increasing precision a NLO-analysis will be the next task since these corrections are expected to be sizeable. See, e.g. \cite{Anikin:2013aka, Anikin:2015ita} for the 
case of the electromagnetic form factors. This is a huge calculation since in contrast to \cite{Anikin:2013aka} there will be an additional mass scale due to the heavy quark and 
additional structures contributing at the same order of the twist expansion. On the other hand combined with experimental data of sufficient precision 
it will give an excellent complementary possibility to make quantitative statements on the low lying nucleon resonances in terms of the fundamental degrees of freedom of QCD.
\section*{Acknowledgements}
We appreciate helpful discussion with V. M. Braun. This work was funded by the BMBF under the contract number 05P12WRFTE.
\appendix
\section{Correlation functions} \label{functions}
The coefficient functions $w_{jn}^{(i)}$ are listed below for:
\paragraph{pseudoscalar interpolating current}
\begin{align*}
 w_{11}^{(\mathcal{P})}&=x_2 m_{N^*} \phi_1^{(\mathcal{P})}, \qquad w_{12}^{(\mathcal{P})} = x_2 m_{N^*}^3 \left[ x_2 \phi_2^{(\mathcal{P})}+2 \phi_3^{(\mathcal{P})} \right] ,\\
 w_{13}^{(\mathcal{P})}&=4 x_2 m_{N^*}^2 m_c^2 \phi_3^{(\mathcal{P})}, \\ \\
 w_{21}^{(\mathcal{P})}&=w_{23}^{(\mathcal{P})}=0, \qquad w_{22}^{(\mathcal{P})}=x_2 m_{N^*}^2 \phi_2^{(\mathcal{P})}, \\ \\
 w_{31}^{(\mathcal{P})}&=\frac{m_{N^*}}{2} (m_c+x_2 m_{N^*}) \phi_1^{(\mathcal{P})}, \\
 w_{32}^{(\mathcal{P})}&=\frac{m_{N^*}^2}{2} \left[ m_c(m_c+x_2m_{N^*})\phi_2^{(\mathcal{P})}+2x_2 m_{N^*}^2 \phi_3^{(\mathcal{P})} \right],\\
 w_{33}^{(\mathcal{P})}&= 2 m_{N^*}^3m_c^2 (m_c+x_2 m_{N^*}) \phi_3^{(\mathcal{P})} ,\\ \\
 w_{41}^{(\mathcal{P})}&=\frac{m_{N^*}}{2} \phi_1^{(\mathcal{P})} , \qquad w_{42}^{(\mathcal{P})}= \frac{m_{N^*}^2}{2} \left[ m_c \phi_2^{(\mathcal{P})} + 2 m_{N^*} \phi_3^{(\mathcal{P})} \right], \\
 w_{43}^{(\mathcal{P})}&=2 m_{N^*}^3 m_c^2 \phi_3^{(\mathcal{P})}, \\ \\
 w_{51}^{(\mathcal{P})}&=-m_{N^*} \phi_1^{(\mathcal{P})}, \qquad w_{52}^{(\mathcal{P})}=-m_{N^*}^3 \left[ x_2 \phi_2^{(\mathcal{P})}+2 \phi_3^{(\mathcal{P})} \right], \\
 w_{53}^{(\mathcal{P})}&=-4 m_{N^*}^3 m_c^2 \phi_3^{(\mathcal{P})},\\ \\
 w_{61}^{(\mathcal{P})}&=w_{63}^{(\mathcal{P})}=0, \qquad w_{62}^{(\mathcal{P})}=-m_{N^*}^2 \phi_2^{(\mathcal{P})},
\end{align*}
where the functions $\phi_i^{\mathcal{P}}$ are
\begin{align*}
 \phi_1^{(\mathcal{P})}=&2 \tilde{A}_1+4 \tilde{A}_3+2\tilde{A}_{123}+2\tilde{P}_1+2 \tilde{S}_1+6 \tilde{T}_1-12 \tilde{T}_7-\tilde{T}_{123}-5 \tilde{T}_{127}-2 \tilde{V}_1+4 \tilde{V}_3+2\tilde{V}_{123} ,\\
 \phi_2^{(\mathcal{P})} =& 3 \tilde{\tilde{A}}_{34}+2 \tilde{\tilde{A}}_{123}-\tilde{\tilde{A}}_{1345}-2\tilde{\tilde{P}}_{21}+2\tilde{\tilde{S}}_{12}-12 \tilde{\tilde{T}}_{78}-2\tilde{\tilde{T}}_{123}-4\tilde{\tilde{T}}_{127}-6 \tilde{\tilde{T}}_{158}+\tilde{\tilde{T}}_{234578} \\
 & -3\tilde{\tilde{V}}_{43}+2 \tilde{\tilde{V}}_{123}+\tilde{\tilde{V}}_{1345}, \\
 \phi_3^{(\mathcal{P})}=&-\tilde{A}_1^{M}-3\tilde{T}_1^M+\tilde{V}_1^M+\tilde{\tilde{\tilde{A}}}_{123456}-3 \tilde{\tilde{\tilde{T}}}_{125678}+\tilde{\tilde{\tilde{T}}}_{234578}+\tilde{\tilde{\tilde{V}}}_{123456}.
\end{align*}

\paragraph{axial-vector interpolating current}
\begin{align*}
 w_{11}^{(\mathcal{A})}&=2\left[-2 m_c \phi_1^{(\mathcal{A})} -x_2 m_{N^*}(2\phi_1^{(\mathcal{A})}+\phi_2^{(\mathcal{A})} ) +2m_{N^*} \phi_3^{(\mathcal{A})}\right],\\
 w_{12}^{(\mathcal{A})}&=2 m_{N^*} \left[ x_2^2 m_{N^*}^2 \phi_4^{(\mathcal{A})} -x_2 m_{N^*}m_c \phi_5^{(\mathcal{A})} +2 m_c^2 \phi_3 ^{(\mathcal{A})}+2x_2 m_{N^*}^2 \phi_6^{(\mathcal{A})} \right], \\
 w_{13}^{(\mathcal{A})}&=8 m_{N^*}^2 m_c \left[-m_c^2 \phi_7^{(\mathcal{A})} + x_2 m_{N^*} m_c \phi_6^{(\mathcal{A})} -x_2^2 m_{N^*}^2 \phi_8^{(\mathcal{A})}\right], \\ \\
 w_{21}^{(\mathcal{A})}&=-4\phi_1^{(\mathcal{A})}, \qquad w_{23}^{(\mathcal{A})}=8 m_{N^*}^2 m_c \left[- m_c \phi_7^{(\mathcal{A})} - x_2 m_{N^*} \phi_8^{(\mathcal{A})} \right], \\
 w_{22}^{(\mathcal{A})}&= 2m_{N^*} \left[2m_c \phi_3^{(\mathcal{A})} +x_2 m_{N^*} \phi_4^{(\mathcal{A})} -2 m_{N^*} \phi_7^{(\mathcal{A})} \right], \\ \\
 w_{31}^{(\mathcal{A})}&=-2\frac{m_c^2-q^2}{ x_2} \phi_1^{(\mathcal{A})} +m_{N^*} m_c \phi_2^{(\mathcal{A})} -m_{N^*}^2 \phi_9^{(\mathcal{A})} +2 x_2 m_{N^*}^2 \phi_{10}^{(\mathcal{A})} +2\frac{m_{N^*}^2}{ x_2} \phi_7^{(\mathcal{A})},\\ 
 w_{32}^{(\mathcal{A})}&=m_{N^*}^2 \left[-2(q^2-x_2^2 m_{N^*}^2)\phi_3^{(\mathcal{A})} +2 \frac{q^2+m_c^2}{ x_2}\phi_7^{(\mathcal{A})} +x_2 m_{N^*}m_c \phi_{11}^{(\mathcal{A})}\right. \\
 &\left.-m_c^2 (\phi_5^{(\mathcal{A})}-2\phi_3^{(\mathcal{A})})  +2m_{N^*}(m_c-x_2 m_{N^*}) \phi_8^{(\mathcal{A})} \right],\\
 w_{33}^{(\mathcal{A})}&=4\frac{m_c^2 m_{N^*}^2}{x_2} \left[ -(m_c^2-q^2) \phi_7^{(\mathcal{A})} +x_2 m_{N^*} m_c \phi_{12}^{(\mathcal{A})} -x_2^2 m_{N^*}^2 \phi_8^{(\mathcal{A})} \right],\\ \\
 w_{41}^{(\mathcal{A})}&=2m_{N^*} \left[  (\phi_1^{(\mathcal{A})}+\phi_{10}^{(\mathcal{A})})- \frac{\phi_3^{(\mathcal{A})}}{x_2}  \right] , \qquad w_{43}^{(\mathcal{A})} =4 m_{N^*}^3 m_c^2 \phi_{13}^{(\mathcal{A})},\\
 w_{42}^{(\mathcal{A})}&=\frac{m_{N^*}}{x_2} \left[ 2(m_c^2+x_2^2 m_{N^*}^2-q^2) \phi_3^{(\mathcal{A})} +x_2 m_{N^*}m_c \phi_{11}^{(\mathcal{A})} +2x_2 m_{N^*}^2  \phi_{13}^{(\mathcal{A})} \right], \\ \\
 w_{51}^{(\mathcal{A})}&=2m_{N^*} \phi_2^{(\mathcal{A})}, \qquad w_{53}^{(\mathcal{A})}= 8m_{N^*}^3 m_c \left[ m_c \phi_{14}^{(\mathcal{A})} +x_2 m_{N^*} \phi_8^{(\mathcal{A})}  \right],\\
 w_{52}^{(\mathcal{A})}&=2m_{N^*}^2 \left[-x_2 m_{N^*} \phi_4^{(\mathcal{A})} +m_c (\phi_5^{(\mathcal{A})}+2\phi_3^{(\mathcal{A})}) +2m_{N^*} \phi_{14}^{(\mathcal{A})}\right], \\ \\
 w_{61}^{(\mathcal{A})}&=0, \qquad w_{62}^{(\mathcal{A})}=-2m_{N^*}^2 \phi_4^{(\mathcal{A})} , \qquad w_{63}^{(\mathcal{A})}=8 m_{N^*}^3 m_c \phi_8^{(\mathcal{A})},
\end{align*}
where the functions $\phi_i^{\mathcal{A}}$ are
\begin{align*}
\phi_1^{(\mathcal{A})} &= \tilde{A_1}+2\tilde{T_1}+\tilde{V_1},\\
\phi_2^{(\mathcal{A})} &= 2\tilde{A_3}-2\tilde{P_1}+2\tilde{S_1}-2\tilde{T_1}+\tilde{T}_{123}+\tilde{T}_{127}-2\tilde{V_3},\\
\phi_3^{(\mathcal{A})} &=  \tilde{\tilde{A}}_{123}-\tilde{\tilde{T}}_{123}-\tilde{\tilde{T}}_{127}-\tilde{\tilde{V}}_{123},\\
\phi_4^{(\mathcal{A})} &=  -\tilde{\tilde{A}}_{34}+\tilde{\tilde{A}}_{1345}-2\tilde{\tilde{P}}_{21}-2\tilde{\tilde{S}}_{12}-2\tilde{\tilde{T}}_{127}+2\tilde{\tilde{T}}_{158}- \tilde{\tilde{T}}_{234578}- \tilde{\tilde{V}}_{43}+ \tilde{\tilde{V}}_{1345},\\
\phi_5^{(\mathcal{A})} &=  -\tilde{\tilde{A}}_{34}-2\tilde{\tilde{A}}_{123}- \tilde{\tilde{A}}_{1345}+4\tilde{\tilde{T}}_{127}-4\tilde{\tilde{T}}_{158}+2\tilde{\tilde{T}}_{234578}-\tilde{\tilde{V}}_{43}+2\tilde{\tilde{V}}_{123}- \tilde{\tilde{V}}_{1345}, \\
\phi_6^{(\mathcal{A})} &= \tilde{A}_1^M+\tilde{T}_1^M+\tilde{V}_1^M-\tilde{\tilde{\tilde{A}}}_{123456}+\tilde{\tilde{\tilde{T}}}_{125678}-\tilde{\tilde{\tilde{T}}}_{234578}+\tilde{\tilde{\tilde{V}}}_{123456},\\
\phi_7^{(\mathcal{A})} &=  -\tilde{A}_1^M-2\tilde{T}_1^M-\tilde{V}_1^M+ \tilde{\tilde{\tilde{T}}}_{234578} ,\\
\phi_8^{(\mathcal{A})} &= \tilde{\tilde{\tilde{A}}}_{123456}-2\tilde{\tilde{\tilde{T}}}_{125678}+\tilde{\tilde{\tilde{T}}}_{234578}-\tilde{\tilde{\tilde{V}}}_{123456},\\
\phi_9^{(\mathcal{A})} &= -2\tilde{\tilde{A}}_{34}-2\tilde{\tilde{A}}_{123}-2\tilde{\tilde{P}}_{21}-2\tilde{\tilde{S}}_{12}+2\tilde{\tilde{T}}_{127}-2\tilde{\tilde{T}}_{158}+\tilde{\tilde{T}}_{234578}-2\tilde{\tilde{V}}_{43}+2\tilde{\tilde{V}}_{123},\\
\phi_{10}^{(\mathcal{A})} &= \tilde{A}_{123}-\tilde{T}_{123}-\tilde{T}_{127}-\tilde{V}_{123},\\
\phi_{11}^{(\mathcal{A})} &= 2\tilde{\tilde{A}}_{34}+2\tilde{\tilde{P}}_{21} +2\tilde{\tilde{S}}_{12}+2\tilde{\tilde{T}}_{123}+2\tilde{\tilde{T}}_{158}- \tilde{\tilde{T}}_{234578}+2\tilde{\tilde{V}}_{43},\\
\phi_{12}^{(\mathcal{A})} &= \tilde{T}_1^M+\tilde{\tilde{\tilde{T}}}_{125678}-\tilde{\tilde{\tilde{T}}}_{234578},\\
\phi_{13}^{(\mathcal{A})} &=  -\tilde{A}_1^M-2\tilde{T}_1^M-\tilde{V}_1^M-\tilde{\tilde{\tilde{A}}}_{123456}+2\tilde{\tilde{\tilde{T}}}_{125678}+\tilde{\tilde{\tilde{V}}}_{123456},\\
\phi_{14}^{(\mathcal{A})} &= \tilde{T}_1^M+\tilde{\tilde{\tilde{A}}}_{123456}-\tilde{\tilde{\tilde{T}}}_{125678}-\tilde{\tilde{\tilde{V}}}_{123456}.
\end{align*}
They agree with those in \cite{Khodjamirian:2011jp} if one replaces $m_N\to m_{N^*}$ and $m_c\to -m_c$.


\begin{thebibliography}{99}
\bibliographystyle{unsrt}

%\cite{Khodjamirian:2011jp}
\bibitem{Khodjamirian:2011jp}
  A.~Khodjamirian, C.~Klein, T.~Mannel and Y.~-M.~Wang,
  %``Form Factors and Strong Couplings of Heavy Baryons from QCD Light-Cone Sum Rules,''
  JHEP {\bf 1109}, 106 (2011).
  %[arXiv:1108.2971 [hep-ph]].
  %%CITATION = ARXIV:1108.2971;%
  
  
  %\cite{Anikin:2015ita}
\bibitem{Anikin:2015ita}
  I.~V.~Anikin, V.~M.~Braun and N.~Offen,
  %``Electroproduction of the $N^*(1535)$ nucleon resonance in QCD,''
  Phys.\ Rev.\ D {\bf 92} (2015) 1,  014018
  [arXiv:1505.05759 [hep-ph]].
  
  
  
  %\cite{Braun:2014wpa}
\bibitem{Braun:2014wpa}
  V.~M.~Braun {\it et al.},
  %``Light-cone Distribution Amplitudes of the Nucleon and Negative Parity Nucleon Resonances from Lattice QCD,''
  Phys.\ Rev.\ D {\bf 89} (2014) 9,  094511
  [arXiv:1403.4189 [hep-lat]].
  
 


  
%\cite{Aznauryan:2012ba}
\bibitem{Aznauryan:2012ba}
  I.~G.~Aznauryan {\it et al.},
  %A.~Bashir, V.~Braun, S.~J.~Brodsky, V.~D.~Burkert, L.~Chang, C.~Chen and B.~El-Bennich {\it et al.},
  %``Studies of Nucleon Resonance Structure in Exclusive Meson Electroproduction,''
  Int.\ J.\ Mod.\ Phys.\ E {\bf 22}, 1330015 (2013).
  %[arXiv:1212.4891 [nucl-th]].
  %%CITATION = ARXIV:1212.4891;%%
R.W.Gothe et al., Nucleon Resonance Studies with CLAS12, Experiment E12-09-003.

\bibitem{LCSR}
%\cite{Balitsky:1986st}
%\bibitem{Balitsky:1986st}
  I.~I.~Balitsky, V.~M.~Braun and A.~V.~Kolesnichenko,
  %``SIGMA+ ---> P GAMMA DECAY IN QCD. (IN RUSSIAN),''
  Sov.\ J.\ Nucl.\ Phys.\  {\bf 44}, 1028 (1986);
  %[Yad.\ Fiz.\  {\bf 44}, 1582 (1986)].
  %%CITATION = SJNCA,44,1028;%%
%\cite{Balitsky:1989ry}
%\bibitem{Balitsky:1989ry}
  %I.~I.~Balitsky, V.~M.~Braun and A.~V.~Kolesnichenko,
  %``Radiative Decay Sigma+ ---> p gamma in Quantum Chromodynamics,''
  Nucl.\ Phys.\ B {\bf 312}, 509 (1989);
  %%CITATION = NUPHA,B312,509;%%
%\cite{Chernyak:1990ag}
%\bibitem{Chernyak:1990ag}
  V.~L.~Chernyak and I.~R.~Zhitnitsky,
  %``B meson exclusive decays into baryons,''
  Nucl.\ Phys.\ B {\bf 345}, 137 (1990).
  %%CITATION = NUPHA,B345,137;%%
%\cite{Shifman:1978bx}
\bibitem{Shifman:1978bx}
  M.~A.~Shifman, A.~I.~Vainshtein and V.~I.~Zakharov,
  %``QCD and Resonance Physics. Theoretical Foundations,''
  Nucl.\ Phys.\ B {\bf 147} (1979) 385.
  doi:10.1016/0550-3213(79)90022-1
  %%CITATION = doi:10.1016/0550-3213(79)90022-1;%%
  %4499 citations counted in INSPIRE as of 10 Dec 2015
  
  
  %\cite{Braun:2009jy}
\bibitem{Braun:2009jy}
  V.~M.~Braun  {\it et al.},
  %``Electroproduction of the N*(1535) resonance at large momentum transfer,''
  Phys.\ Rev.\ Lett.\  {\bf 103}, 072001 (2009).
  %[arXiv:0902.3087 [hep-ph]].
  %%CITATION = ARXIV:0902.3087;%%
  

  
    %\cite{Anikin:2013aka}
\bibitem{Anikin:2013aka}
  I.~V.~Anikin, V.~M.~Braun and N.~Offen,
  %``Nucleon Form Factors and Distribution Amplitudes in QCD,''
  Phys.\ Rev.\ D {\bf 88} (2013) 114021
  doi:10.1103/PhysRevD.88.114021
  [arXiv:1310.1375 [hep-ph]].
  %%CITATION = doi:10.1103/PhysRevD.88.114021;%%
  %12 citations counted in INSPIRE as of 10 Dec 2015


  %%%%%%%%% Hadronic decay review %%%%%%%%%%%%%%%%%%%%
  
  %\cite{Oset:2016lyh}
\bibitem{Oset:2016lyh}
  E.~Oset {\it et al.},
  %``Weak decays of heavy hadrons into dynamically generated resonances,''
  Int.\ J.\ Mod.\ Phys.\ E {\bf 25} (2016) no.01,  1630001
  doi:10.1142/S0218301316300010
  [arXiv:1601.03972 [hep-ph]].
  %%CITATION = doi:10.1142/S0218301316300010;%%
  %6 citations counted in INSPIRE as of 18 Apr 2016
  
%%%%%%%%%%%%%%% Lambda_b decays in LCSR %%%%%%%%%%%%%%%%%%%%%%%%%%%%%%%%
  
  %\cite{Huang:2004vf}
\bibitem{Huang:2004vf}
  M.~-q.~Huang and D.~-W.~Wang,
  %``Light cone QCD sum rules for the semileptonic decay Lambda(b) -> p l anti-nu,''
  Phys.\ Rev.\ D {\bf 69}, 094003 (2004).
  %[hep-ph/0401094].
  %%CITATION = HEP-PH/0401094;%%

%\cite{Wang:2009hra}
\bibitem{Wang:2009hra}
  Y.~-M.~Wang, Y.~-L.~Shen and C.~-D.~Lu,
  %``Lambda(b) ---> p, Lambda transition form factors from QCD light-cone sum rules,''
  Phys.\ Rev.\ D {\bf 80}, 074012 (2009).
  %[arXiv:0907.4008 [hep-ph]].
  %%CITATION = ARXIV:0907.4008;%%

%\cite{Aliev:2010uy}
%\bibitem{Aliev:2010uy}
%  T.~M.~Aliev, K.~Azizi and M.~Savci,
  %``Analysis of the $Lambda_{b}\rightarrow \Lambda \ell^+\ell^- $ decay in QCD,''
%  Phys.\ Rev.\ D {\bf 81}, 056006 (2010)
  %[arXiv:1001.0227 [hep-ph]].
  %%CITATION = ARXIV:1001.0227;%%


  
\bibitem{Azizi:2009wn}
  K.~Azizi, M.~Bayar, Y.~Sarac and H.~Sundu,
  %``Semileptonic Lambda(b,c) to Nucleon Transitions in Full QCD at Light Cone,''
  Phys.\ Rev.\ D {\bf 80} (2009) 096007
  doi:10.1103/PhysRevD.80.096007
  [arXiv:0908.1758 [hep-ph]].
  %%CITATION = doi:10.1103/PhysRevD.80.096007;%%


%%%%%%%%%%%%%%%%%% Nucleon DAs %%%%%%%%%%%%%%%%%%%%%%%%%%%%%%%%%%%
%\cite{Anikin:2013yoa}
\bibitem{Anikin:2013yoa}
  I.~V.~Anikin and A.~N.~Manashov,
  %``Higher twist nucleon distribution amplitudes in Wandzura-Wilczek approximation,''
  Phys.\ Rev.\ D {\bf 89} (2014) no.1,  014011
  doi:10.1103/PhysRevD.89.014011
  [arXiv:1311.3584 [hep-ph]].
  %%CITATION = doi:10.1103/PhysRevD.89.014011;%%
  %3 citations counted in INSPIRE as of 20 Apr 2016lyh
  
  %\cite{Anikin:2015qos}
\bibitem{Anikin:2015qos}
  I.~V.~Anikin and A.~N.~Manashov,
  %``Baryon octet distribution amplitudes in Wandzura-Wilczek approximation,''
  Phys.\ Rev.\ D {\bf 93} (2016) no.3,  034024
  doi:10.1103/PhysRevD.93.034024
  [arXiv:1512.07141 [hep-ph]].
  %%CITATION = doi:10.1103/PhysRevD.93.034024;%%
%\cite{Braun:2000kw}
\bibitem{Braun:2000kw}
  V.~Braun, R.~J.~Fries, N.~Mahnke and E.~Stein,
  %``Higher twist distribution amplitudes of the nucleon in QCD,''
  Nucl.\ Phys.\ B {\bf 589} (2000) 381
   [Nucl.\ Phys.\ B {\bf 607} (2001) 433]
  doi:10.1016/S0550-3213(00)00516-2
  [hep-ph/0007279].
  %%CITATION = doi:10.1016/S0550-3213(00)00516-2;%%
  %123 citations counted in INSPIRE as of 14 Dec 2015

  %\cite{Braun:2006hz}
\bibitem{Braun:2006hz}
  V.~M.~Braun, A.~Lenz and M.~Wittmann,
  %``Nucleon Form Factors in QCD,''
  Phys.\ Rev.\ D {\bf 73} (2006) 094019
  doi:10.1103/PhysRevD.73.094019
  [hep-ph/0604050].
  %%CITATION = doi:10.1103/PhysRevD.73.094019;%%
  %110 citations counted in INSPIRE as of 22 Dec 2015
%\cite{Braun:2008ia}
\bibitem{Braun:2008ia}
  V.~M.~Braun, A.~N.~Manashov and J.~Rohrwild,
  %``Baryon Operators of Higher Twist in QCD and Nucleon Distribution Amplitudes,''
  Nucl.\ Phys.\ B {\bf 807} (2009) 89
  doi:10.1016/j.nuclphysb.2008.08.012
  [arXiv:0806.2531 [hep-ph]].
  %%CITATION = doi:10.1016/j.nuclphysb.2008.08.012;%%
  %36 citations counted in INSPIRE as of 22 Dec 2015

  
  %%%%%%%%%%%%%%%%% mb und mc %%%%%%%%%%%%%%%%%%%%
  
  %\cite{Chetyrkin:2009fv}
\bibitem{Chetyrkin:2009fv}
  K.~G.~Chetyrkin, J.~H.~Kuhn, A.~Maier, P.~Maierhofer, P.~Marquard, M.~Steinhauser and C.~Sturm,
  %``Charm and Bottom Quark Masses: An Update,''
  Phys.\ Rev.\ D {\bf 80} (2009) 074010
  doi:10.1103/PhysRevD.80.074010
  [arXiv:0907.2110 [hep-ph]].
  %%CITATION = doi:10.1103/PhysRevD.80.074010;%%
  %180 citations counted in INSPIRE as of 22 Dec 2015

  
  %%%%%%%%%%%%%%%%%% mB* %%%%%%%%%%%%%%%%%%%%%%%%%%
  
  %\cite{Wang:2010fq}
\bibitem{Wang:2010fq}
  Z.~G.~Wang,
  %``Analysis of the ${1\over 2}^{\pm}$ antitriplet heavy baryon states with QCD sum rules,''
  Eur.\ Phys.\ J.\ C {\bf 68} (2010) 479
  doi:10.1140/epjc/s10052-010-1365-8
  [arXiv:1001.1652 [hep-ph]].
  %%CITATION = doi:10.1140/epjc/s10052-010-1365-8;%%
  %15 citations counted in INSPIRE as of 22 Dec 2015

  


 %%%%%%%%%%%%%%%%% PDG %%%%%%%%%%%%%%%%%
 
 %\cite{Agashe:2014kda}
\bibitem{Agashe:2014kda}
  K.~A.~Olive {\it et al.} [Particle Data Group Collaboration],
  %``Review of Particle Physics,''
  Chin.\ Phys.\ C {\bf 38} (2014) 090001.
  doi:10.1088/1674-1137/38/9/090001
  %%CITATION = doi:10.1088/1674-1137/38/9/090001;%%
  %2545 citations counted in INSPIRE as of 18 Dec 2015

  
  
%%%%%%%%%%%%%%%%%% Two point sum rules %%%%%%%%%%%%%%%

%\cite{Bagan:1993ii}
\bibitem{Bagan:1993ii}
  E.~Bagan, M.~Chabab, H.~G.~Dosch and S.~Narison,
  %``Baryon sum rules in the heavy quark effective theory,''
  Phys.\ Lett.\ B {\bf 301} (1993) 243.
  doi:10.1016/0370-2693(93)90696-F
  %%CITATION = doi:10.1016/0370-2693(93)90696-F;%%
  %59 citations counted in INSPIRE as of 07 Mar 2016

\bibitem{Yuming}
Private communication with the authors of \cite{Khodjamirian:2011jp}
  
%%%%%%%%%% Form Factor Parametrization %%%%%%%%%%%%%%%

%\cite{Bourrely:2008za}
\bibitem{Bourrely:2008za}
  C.~Bourrely, I.~Caprini and L.~Lellouch,
  %``Model-independent description of B ---> pi l nu decays and a determination of |V(ub)|,''
  Phys.\ Rev.\ D {\bf 79} (2009) 013008
   [Phys.\ Rev.\ D {\bf 82} (2010) 099902]
  doi:10.1103/PhysRevD.82.099902, 10.1103/PhysRevD.79.013008
  [arXiv:0807.2722 [hep-ph]].
  %%CITATION = doi:10.1103/PhysRevD.82.099902, 10.1103/PhysRevD.79.013008;%%
  %122 citations counted in INSPIRE as of 10 Mar 2016

  
  
  
  
  
%%%%%%%%% N* experimental data %%%%%%%%%%%%%%%%%
  
  
 %\cite{Aznauryan:2009mx}
\bibitem{Aznauryan:2009mx}
  I.~G.~Aznauryan {\it et al.}  [CLAS Collaboration],
  %``Electroexcitation of nucleon resonances from CLAS data on single pion electroproduction,''
  Phys.\ Rev.\ C {\bf 80}, 055203 (2009).
  %[arXiv:0909.2349 [nucl-ex]].
  %%CITATION = ARXIV:0909.2349;%%

  
  
%\cite{Denizli:2007tq}
\bibitem{Denizli:2007tq}
  H.~Denizli {\it et al.}  [CLAS Collaboration],
  %``Q*2 dependence of the S(11)(1535) photocoupling and evidence for a P-wave resonance in eta electroproduction,''
  Phys.\ Rev.\ C {\bf 76}, 015204 (2007).
  %[arXiv:0704.2546 [nucl-ex]].
  %%CITATION = ARXIV:0704.2546;%%

%\cite{Dalton:2008aa}
\bibitem{Dalton:2008aa}
  M.~M.~Dalton, G.~S.~Adams, A.~Ahmidouch, T.~Angelescu, J.~Arrington, R.~Asaturyan, O.~K.~Baker and N.~Benmouna {\it et al.},
  %``Electroproduction of Eta Mesons in the S(11)(1535) Resonance Region at High Momentum Transfer,''
  Phys.\ Rev.\ C {\bf 80}, 015205 (2009).
  %[arXiv:0804.3509 [hep-ex]].
  %%CITATION = ARXIV:0804.3509;%%

%\cite{Armstrong:1998wg}
\bibitem{Armstrong:1998wg}
  C.~S.~Armstrong {\it et al.}  [Jefferson Lab E94014 Collaboration],
  %``Electroproduction of the S(11)(1535) resonance at high momentum transfer,''
  Phys.\ Rev.\ D {\bf 60}, 052004 (1999).
  %[nucl-ex/9811001].
  %%CITATION = NUCL-EX/9811001;%%
  
  
  

\end{thebibliography}
\end{document}